\documentclass[aps,prb,twocolumn,superscriptaddress]{revtex4-2}
\usepackage{graphicx,color}

\begin{document}


\title{Absence of hexagonal to square structural transition in LiFeAs vortex matter}



\author{S. Hoffmann}
\affiliation{Leibniz Institute for Solid State and Materials Research, Helmholtzstra{\ss}e 20, 01069 Dresden, Germany} 
\affiliation{Bergische Univertsität Wuppertal, 42119 Wuppertal, Germany }

\author{R. Schlegel}
\affiliation{Leibniz Institute for Solid State and Materials Research, Helmholtzstra{\ss}e 20, 01069 Dresden, Germany}

\author{C. Salazar}
\affiliation{Leibniz Institute for Solid State and Materials Research, Helmholtzstra{\ss}e 20, 01069 Dresden, Germany}

\author{S. Sykora}
\affiliation{Institute of Theoretical Physics, Technische Universit{\"a}t Dresden, 01069 Dresden,Germany}

\author{P. K. Nag}
\affiliation{Leibniz Institute for Solid State and Materials Research, Helmholtzstra{\ss}e 20, 01069 Dresden, Germany}

\author{P. Khanenko}
\affiliation{Max-Planck-Institute for Chemical Physics of Solids, 001187 Dresden, Germany}

\author{R. Beck}
\affiliation{Leibniz Institute for Solid State and Materials Research, Helmholtzstra{\ss}e 20, 01069 Dresden, Germany}

\author{S. Aswartham}
\affiliation{Leibniz Institute for Solid State and Materials Research, Helmholtzstra{\ss}e 20, 01069 Dresden, Germany}

\author{S. Wurmehl}
\affiliation{Leibniz Institute for Solid State and Materials Research, Helmholtzstra{\ss}e 20, 01069 Dresden, Germany}
\affiliation{Institute of Solid State Physics, Technische Universit{\"a}t Dresden, 01069 Dresden,Germany}

\author{B. B\"uchner}
\affiliation{Leibniz Institute for Solid State and Materials Research, Helmholtzstra{\ss}e 20, 01069 Dresden, Germany}
\affiliation{Institute of Solid State Physics, Technische Universit{\"a}t Dresden, 01069 Dresden,Germany}
\affiliation{Center for Transport and Devices, Technische Universit{\"a}t Dresden, 01069 Dresden, Germany}

\author{Y. Fasano}
\affiliation{Leibniz Institute for Solid State and Materials Research, Helmholtzstra{\ss}e 20, 01069 Dresden, Germany}
\affiliation{Centro  Atómico  Bariloche  and  Instituto  Balseiro,  CNEA, CONICET  and  Universidad  Nacional  de  Cuyo,  8400  San  Carlos  de  Bariloche,  Argentina
}

\author{C. Hess}
\affiliation{Leibniz Institute for Solid State and Materials Research, Helmholtzstra{\ss}e 20, 01069 Dresden, Germany}
\affiliation{Bergische Univertsität Wuppertal, 42119 Wuppertal, Germany }


\date{\today}

\begin{abstract}

We investigated magnetic vortices in two stoichiometric LiFeAs samples by means of scanning tunneling microscopy and spectroscopy.  The vortices were revealed by measuring the local electronic density of states (LDOS) at zero bias conductance of samples in magnetic fields between 0.5  and 12 T. From single vortex spectroscopy we extract the Ginzburg-Landau coherence length of both samples as $4.4\pm0.5$ nm and $4.1\pm0.5$~nm, in accordance with previous findings. However, in contrast to previous reports, our study reveals that the reported hexagonal to square-like vortex lattice transition is absent up to 12 T both in field-cooling and zero-field-cooling processes. Remarkably, a highly ordered zero field cooled hexagonal vortex lattice is observed up to 8 T. We argue that several factors are likely to determine the structure of the vortex lattice in LiFeAs such as (i) details of the cooling procedure (ii) sample stoichiometry that alters the formation of nematic fluctuations, (iii) details of the order parameter and (iv) magnetoelastic coupling.

\end{abstract}


\maketitle

\section{Introduction}

The discovery of topological states in iron based superconductors (IBS) \cite{Zhang182,Zhang2019} has recently led to renewed interest in this material  class. In particular this concerns vortex matter where scanning tunneling microscopy (STM) and spectroscopy (STS) \cite{Wang333, Kong2021, Fan2021} provide evidence for Majorana bound states at the cores of magnetic vortices which are considered as a promising platform for quantum computing. In addition, one can expect the analysis of vortices in IBS \cite{Abrikosov1957} to contribute to the understanding of their microscopic superconducting properties \cite{Hoffman2011, Hanaguri2012, Yin2009, Shan2011, Auslaender2009, Suderow2014}.

An interesting candidate for such investigations is LiFeAs because angle-resolved photoemission spectroscopy (ARPES) in combination with density functional theory calculations (DFT) suggest the existence of topological insulating as well as topological Dirac semimetal bands in LiFeAs \cite{Zhang2019}. Further interest in this material is connected to the fact that it  profoundly differs from other IBS since its fermiology seems to be far away from Fermi surface nesting and from an antiferromagnetic instability \cite{Borisenko2010, Kordyuk2011, Umezawa2012, Borisenko2012, Knolle2012, Hess2013, Zeng2013, Pitcher2010, Aswartham2011, Wright2013}. This has led to an ongoing debate about the nature of superconductivity in LiFeAs. From the experimental perspective this material is well suited for surface sensitive techniques such as STM/STS due to its charge neutral surfaces \cite{Lankau2010}.

Previous reports on the vortex matter in LiFeAs showed a vortex lattice, which can be disordered by pinning effects leading to a transition from 6-fold to 4-fold symmetry at high fields. The symmetry transition is accompanied by a locking of the vortex lattice directions to those of the Fe-lattice \cite{Hanaguri2012,Zhang2019a}. However, while Ref.~\onlinecite{Hanaguri2012} observes this transition at magnetic fields around 8 T in their field cooled sample, Ref.~\onlinecite{Zhang2019a} reports a transition at 3-4 T in a zero field cooled sample.

Motivated by the above considerations we use STM and STS to study magnetic vortices in LiFeAs on two different samples using field cooling (FC) and zero field cooling (ZFC) processes.  From a single vortex spectroscopic analysis we estimate the Ginzburg-Landau-coherence length of both samples using a simple model derived from Ginzburg-Landau theory in cylindrical boundary conditions as $\xi_{GL}^{(1)}=4.4 \pm 0.5$ nm and $\xi_{GL}^{(2)} = 4.1 \pm 0.5$ nm, respectively.  Furthermore, our data allow us to investigate the nucleation of vortex matter, as well as the evolution of disorder in the vortex lattice depending on external magnetic fields and field cooling or zero field cooling processes. For the FC sample we observe a highly disordered vortex lattice up to 12 T. In contrast, the ZFC case shows a strongly ordered lattice up to 8 T. In both cases no clear sign of a transition to a fourfold symmetric vortex lattice has been observed, contradicting previous findings \cite{Hanaguri2012,Zhang2019a}. This suggests that, in addition to the cooling process, other, possibly stoichiometry dependent, properties such as details of the superconducting order parameter, nematic fluctuations or magnetoelestic coupling might influence vortex matter in LiFeAs.

\section{Experiment}

Stoichiometric LiFeAs single crystals were grown via the self-flux method, as described in Ref.~\onlinecite{Morozov2010}. Due to their air sensitivity, the samples were mounted to the microscopes inside a glove box with inert Ar atmosphere.  Sample 1 was investigated using a home-built device based on a dip-stick design, which is suitable for measurements from 5 K to room temperature in a 12 T magnet cryostat \cite{Schlegel2014}. For sample 2, a home-built low-temperature STM \cite{Salazar2018} with a base temperature of 300 mK and a maximum field of 9 T was used. The energy resolution of each system is influenced by the measuring temperature and electronic noise. The resolution limits for the dip-stick and 300 mK systems were determined as approximately 0.5 and 0.15 meV, respectively.  Electrochemical etching was employed to prepare tungsten (W) tips that were used for all measurements. The samples were cleaved in ultra high vacuum conditions with the purpose to obtain flat and clean surfaces suited for STM. The vortex lattice of sample 1 and 2 was studied in field cooling (FC) and zero field cooling (ZFC) conditions, respectively. Maps of the differential conductance dI/dU as a measure of the local density of state (LDOS) were acquired at zero bias in order to reveal the vortices. Individual  spectra in the range [-15, 15] mV have been recorded at selected positions.

\section{Results}

\subsection{Single vortex analysis}

\begin{figure}[ht]
\includegraphics[scale=0.4]{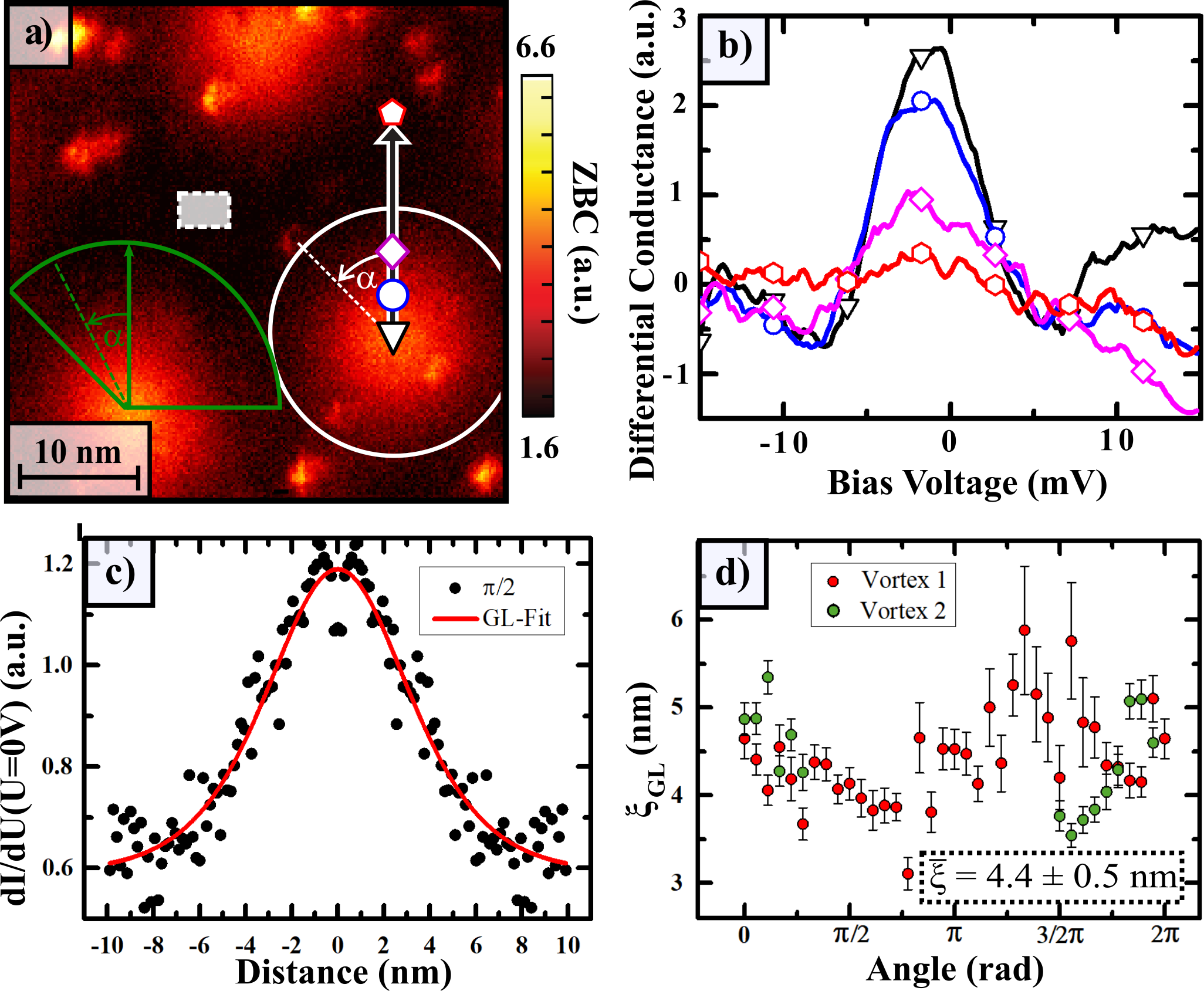}
\caption{(a) ZBC-map of 40 x 40 nm$^2$ for sample 1 measured under FC conditions at $B=6$~T and $T=5$~K. The white and green circumference shows the area where the coherence length values were calculated; these values are shown in (d) as a function of angle. The gray box shows a place where spectra without vortex influence were taken. The average was used to normalize all single point spectra in (b), recorded at positions indicated in (a). (c) ZBC along the arrow in (a) mirrored around the vortex center. The coherence length fit is depicted with the red line. (d) Calculated coherence length along the white circumference in (a). The calculated average value is 4.4 $\pm$ 0.5 nm.}
\label{fig:4_single-vortex-analysis}
\end{figure}

In FIG.~\ref{fig:4_single-vortex-analysis}(a), showing the zero bias conductance (ZBC) of a region of sample 1 recorded at 5 K, one can clearly recognize regions of enhanced ZBC indicative of three vortices. The geometrical symbols and the arrow indicate the place where the spectra depicted in (b) and the ZBC profile presented in (c) were taken. In order to highlight the change of LDOS inside vortices, the spatially averaged undisturbed LDOS (recorded at the gray, dashed-lined area) was subtracted from all spectra shown in FIG.~\ref{fig:4_single-vortex-analysis}(b). The spectra show typical bound states, which are recognizable by an increment of the LDOS around the Fermi level ($E_F$) as well as a reduction of the DOS at the position of the main coherence peaks at about $\pm$6 mV. The form of the spectra corresponds to the theoretical expectation of bound states at the vortex core \cite{Shore1989, Gygi1991, Hayashi1996}, although the details of the inner structure of the vortex is smeared out due to the relatively high temperature (5 K). The DOS shows a characteristic asymmetric distribution around ($E_{F}$), which has its maximum at about -1.2 mV. The latter is consistent with the observations already reported in \cite{Hanaguri2012}, where the maximum of the peak appears at about -0.9 mV.

We use the spatial evolution of the bound states to extract the Ginzburg-Landau coherence length $\xi_{GL}$. In order to justify our approach and to go beyond the commonly used phenomenological fit of an exponential decay \cite{Zhang2019a, Yin2009,Shan2011} we employ Ginzburg-Landau theory in cylindrical boundary conditions. More specifically, we consider a vortex as a microscopic disturbance of the superconductor realizing a quantum well with bound states \cite{Gygi1991,Hess1989}. These states, which are located in the center of a vortex core, can be detected by STM/STS as shown before \cite{Hanaguri2012}. The electrons forming the bound states can be considered normal conducting \cite{Gygi1991} due to the pair breaking nature of the magnetic field inside the vortex core. Note that our model is still valid at the considered temperature since a possible thermal broadening only affects spectral properties of the bound sates but does not influence the spatial decay. In order to describe the spatial evolution of the superconducting and normal states in a vortex core under these conditions, we consider a system with a constant number of electrons, where one part of the electrons is paired and belongs to the superconducting state. The total number of electrons is controlled through the Fermi level. Considering the volume and the number of particles, it is possible to define a general wave function ($\psi_{total}$) in which all electrons are represented by the total density of particles ($|\psi_{total}(r)|^2$). The total density of particles of the vortex system can be expressed as


\begin{equation}
|\psi_{total}(r)|^2 = |\psi_{N}(r)|^2 + |\psi_{SC}(r)|^2.
\label{eq:Psitotal}
\end{equation}

\noindent

Here, $\psi_N(r)$ and $\psi_{SC}(r)$ are the normal and superconducting wave functions, respectively. Qualitatively, we expect that the amplitude of the superconducting wave function $\psi_{SC}$ in the area of a vortex is reduced until its value reaches zero at the vortex center. Vice versa, within a vortex, the number of normal electrons increases. The normal conducting state that is present within the vortex can be described with the wave function $\psi_{N}$, which in contrast to $\psi_{SC}$ has a maximum amplitude at the vortex center and vanishes outside of the vortex \cite{Meissner1960}. We approximate the normal region in the vortex core with a cylindrical area and use  Ginzburg-Landau theory for describing the spatial decay of $\psi_{SC}$ yielding  \cite{Tinkham}

\begin{equation}
|\psi_{SC}(r)|= |\psi_{\infty}| \tanh\left(\frac{r}{\xi_{GL}}\right).
\end{equation}

\noindent

We now identify $\psi_{N}(E_F,r)^2 \propto \text{LDOS}(U=0)$ and thus $dI/dU$. We hence have

\begin{equation}
\frac{dI}{dU}(U=0, r)\propto |\psi_{total}(r)|^2 - |\psi_{\infty}|^2 \tanh^2\left(\frac{r}{\xi_{GL}}\right).
\end{equation}

\noindent
 
In order to apply this result to our data we use that $|\psi_{total}(r)|^2$ is a constant and find

\begin{equation}
\frac{dI}{dU}(U=0, r)\propto A - B\tanh^2\left(\frac{x}{\xi_{GL}}\right),
\end{equation}

\noindent

where A and B are constants and $\xi_{GL}$ is the relevant parameter to determine.

\begin{figure}[ht]
\includegraphics[scale=0.4]{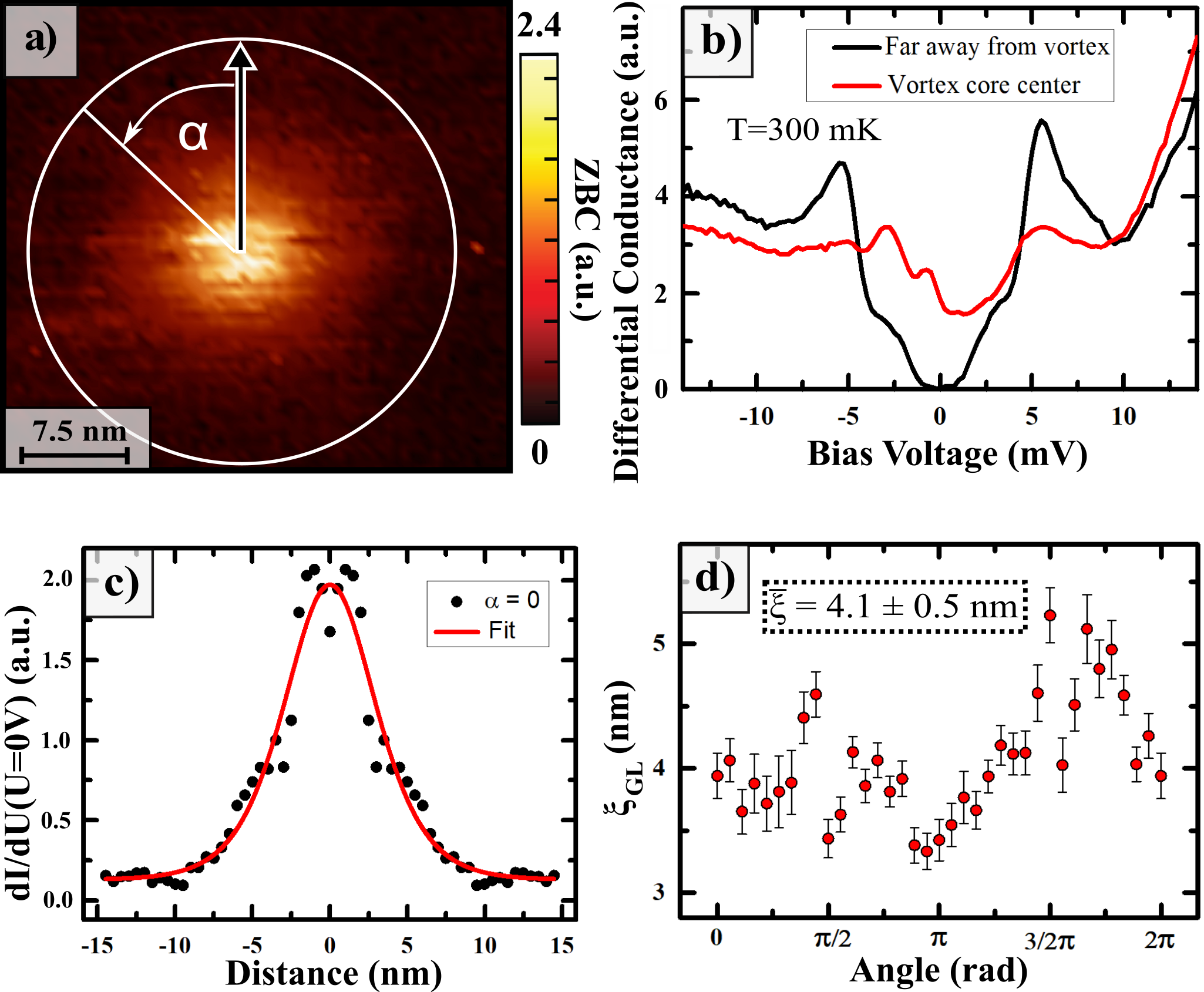}
\caption{(a) ZBC-map of 35 x 35 nm$^2$ for sample 2 measured under ZFC conditions at $B=0.5$~T and $T=300$~mK. The white circumference shows the area where the coherence length values were calculated; these values are shown in (d)  as a function of the angle $\alpha$ which is defined in (a). (b) Single point spectra for sample 2 at a place far away from the vortex and in the center of the vortex, respectively  ($T=$300~mK).  (c) ZBC along the arrow in (a) mirrored around the vortex center. The coherence length fit is depicted with the red line. (d) Calculated coherence length along the white circumference in (a). The calculated average value is 4.1 $\pm$ 0.5 nm.} 
\label{fig:5_single-vortex-analysis}
\end{figure}

FIG.~\ref{fig:4_single-vortex-analysis}(c) shows the ZBC values along the arrow starting from the center of the vortex, as indicated in (a), plotted over the distance. The vortex core center was determined by Gaussian fits as described in the appendix. In order to ensure better fitting results we mirrored the data at its origin.  The corresponding fit is illustrated in FIG.~\ref{fig:4_single-vortex-analysis}(c) by the red line, yielding the Ginzburg-Landau coherence length. Apparently, the spatial decay of the ZBC in (c) can be very well described with equation 4. By rotating the arrow by 360° and performing the fit as described above in regular intervals, we can plot the determined values of $\xi_{GL}$ as a function of the angle $\alpha$ as is shown in FIG.~\ref{fig:4_single-vortex-analysis}(d).  The same procedure was repeated for the second vortex in the image frame. Due to its positioning at the edge of the field of view only an area associated to an arc of 140°  was analyzed (indicated by the green circular segment).  The analysis of both vortex cores resulted in a mean value for the Ginzburg-Landau coherence length of $\xi_{GL}^{(1)}=4.4 \pm 0.5$ nm. Note, that we discarded data points which originate from sites with defect enhanced LDOS (visible as bright spots in FIG.~\ref{fig:4_single-vortex-analysis}(a)). However, defect bound states are known to decay over multiple nanometers and can vary in intensity \cite{Grothe2012}. It is therefore difficult to completely mitigate their influence on the analysis. We believe such defect bound states to be the main reason for the increase in $\xi_{GL}(\alpha)$ at certain angles as is apparent from FIG.~\ref{fig:4_single-vortex-analysis},\ref{fig:5_single-vortex-analysis} and \ref{fig:SingleVortexsample2Map}. This is supported by the error bars increasing with $\xi_{GL}(\alpha)$ indicating, that dI/dU(U=0V,r) is deviating from the expected $tanh^2$ behavior at these angles. In addition FIG.~\ref{fig:SingleVortexsample2Map_Mean} reveals that on average a general low-symmetric anisotropy is present in the data which we attribute to a possible influence of drift (see appendix).

The corresponding results for sample 2 measured with higher energy resolution are presented in FIG.~\ref{fig:5_single-vortex-analysis}. Panel (a) of FIG.~\ref{fig:5_single-vortex-analysis} shows a ZBC map of a single vortex core at a magnetic field of 0.5 T and a temperature of 300~mK. The lower temperature allows for higher resolution single point spectroscopy to be performed. In FIG.~\ref{fig:5_single-vortex-analysis}(b) spectra recorded at points far away from a vortex (black) and at its center (red) are shown. Far away, we observe the previously reported \cite{Hanaguri2012,Grothe2012} double gap structure of LiFeAs. Inside the vortex, an apparent vortex bound state can be identified through the peak at $U_{bias}\approx -0.9$ mV, again matching the findings of Ref.~\onlinecite{Hanaguri2012}.

The coherence length analysis was performed by plotting the ZBC values over the distance from the vortex core center and fitting the data using our model (FIG.~\ref{fig:5_single-vortex-analysis}(c)). This was repeated for multiple angles in the full circumference of the vortex as is marked in FIG.~\ref{fig:5_single-vortex-analysis} (a) by the white circle. The resulting values for $\xi_{GL}(\alpha)$ are plotted in FIG.~\ref{fig:5_single-vortex-analysis} (d).  By calculating the average we obtain a Ginzburg-Landau coherence length of $\xi_{GL}^{(2)} = 4.1 \pm 0.5$ nm for sample 2. Additionally, 5 other vortex cores in sample 2 were analyzed using a larger ZBC map at $B= 2 $~T and $T=6 $~K. The same value for $\xi_{GL}$ was reached within the error for all studied vorticies (see FIG.~\ref{fig:5_single-vortex-analysis} of appendix). Thus, the analysis of the coherence length of sample 1 yields the same value within error bars and is in accordance with reports from literature \cite{Song2010,Zhang2011,Cho2011,Khim2011,Inosov2010,Kurita2011,Lee2009,Li2013,Heyer2011}.

\subsection{Vortex lattice analysis}

\begin{figure*}
\includegraphics[scale=0.61]{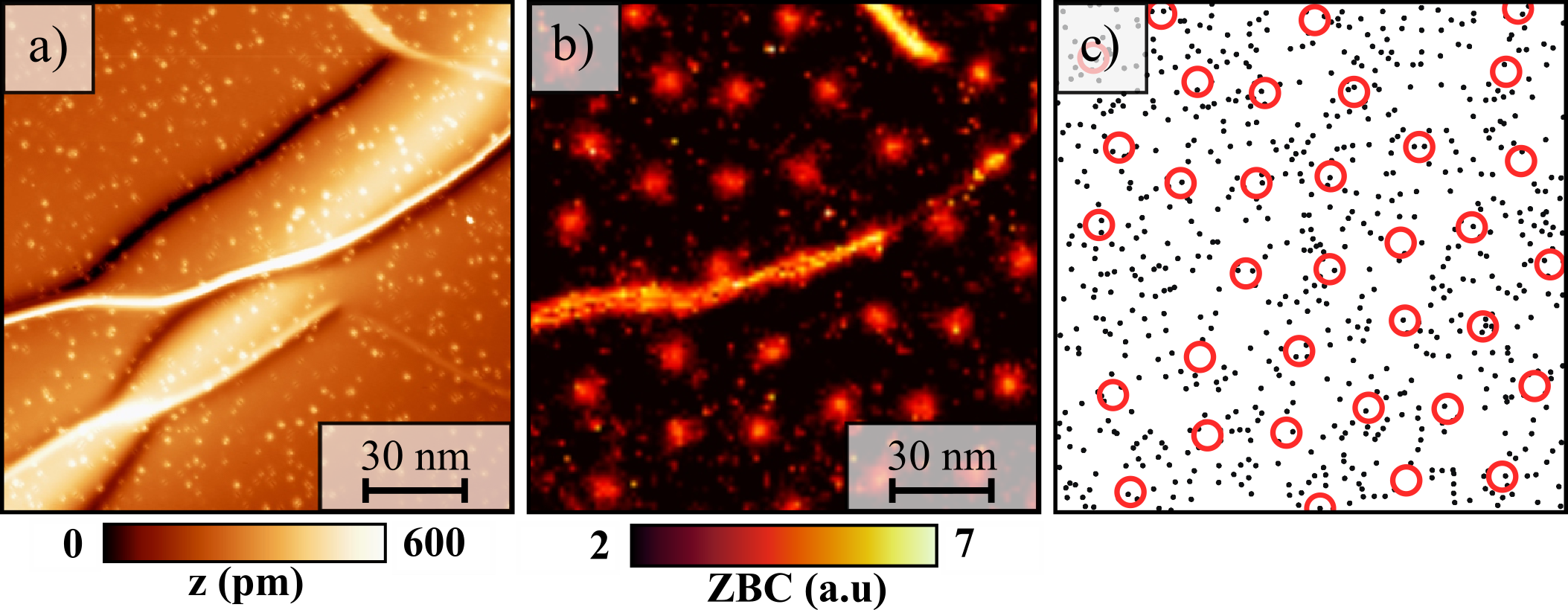}
\caption{(a) Topographic image (150 nm x 150 nm) of the sample showing different atomic and line defects.   ($U = -35$~mV, $I = 300 $~pA, $T = 5$~K), (b) ZBC-map taken in the same area in (a) revealing magnetic vortices as bright spots. The lattice was nucleated at 3 T after following a FC process. (c) s (c)  Positions of the atomic defects (black dots) and vortices (red circles)}
\label{fig:1_vortex_defects}
\end{figure*} 

FIG.~\ref{fig:1_vortex_defects}(a) shows a topographic image of sample 1 where atomic \cite{Schlegel2014, Grothe2012} (see FIG.~\ref{fig:AR}) and line-type defects \cite{Cao2021} are visible. The unfortunate lack of atomically resolved images of these line defects complicates the identification of such structures. Upon closer inspection of FIG.\ref{fig:1_vortex_defects}, however, one notices that the commonly found atomic defects remain visible on top of line defects, indicating an uninterrupted albeit deformed surface layer. In addition, it should be noted that these structure were only observed after sample 1 was cleaved again to clean the surface. It is therefore highly likely that the sample surface was subjected to an unusual amount of force during cleaving, causing the surface to buckle and thereby creating the observed line defects in the form of wrinkles \cite{Cao2021}.   FIG.~\ref{fig:1_vortex_defects}(b) presents an image of the zero bias conductance (ZBC) under the presence of magnetic field (3 T) taken in the same area as in FIG.~\ref{fig:1_vortex_defects}(a) following a FC process. Line defects as well as atomic defects are also recognizable in the ZBC image. Some vortices are directly located on line or atomic defects, others are shifted away from the defects. Since the atomic defects in FIG.~\ref{fig:1_vortex_defects}(b) are not easily recognizable, we highlight them in FIG.~\ref{fig:1_vortex_defects}(c) to allow for a better differentiation between the vortex (red circles) and defect (black dots) positions. A statistic evaluation leads to a number of defects per vortex of 2 $\pm$ 1. The sample clearly shows the presence of a pinning effect on the vortex lattice, which is recognized by the apparently not perfect triangular lattice. However, it is not possible to observe a clear correlation between the surface defects (atomic and line-type) and the vortices.

\begin{figure*}
\includegraphics[scale=0.6]{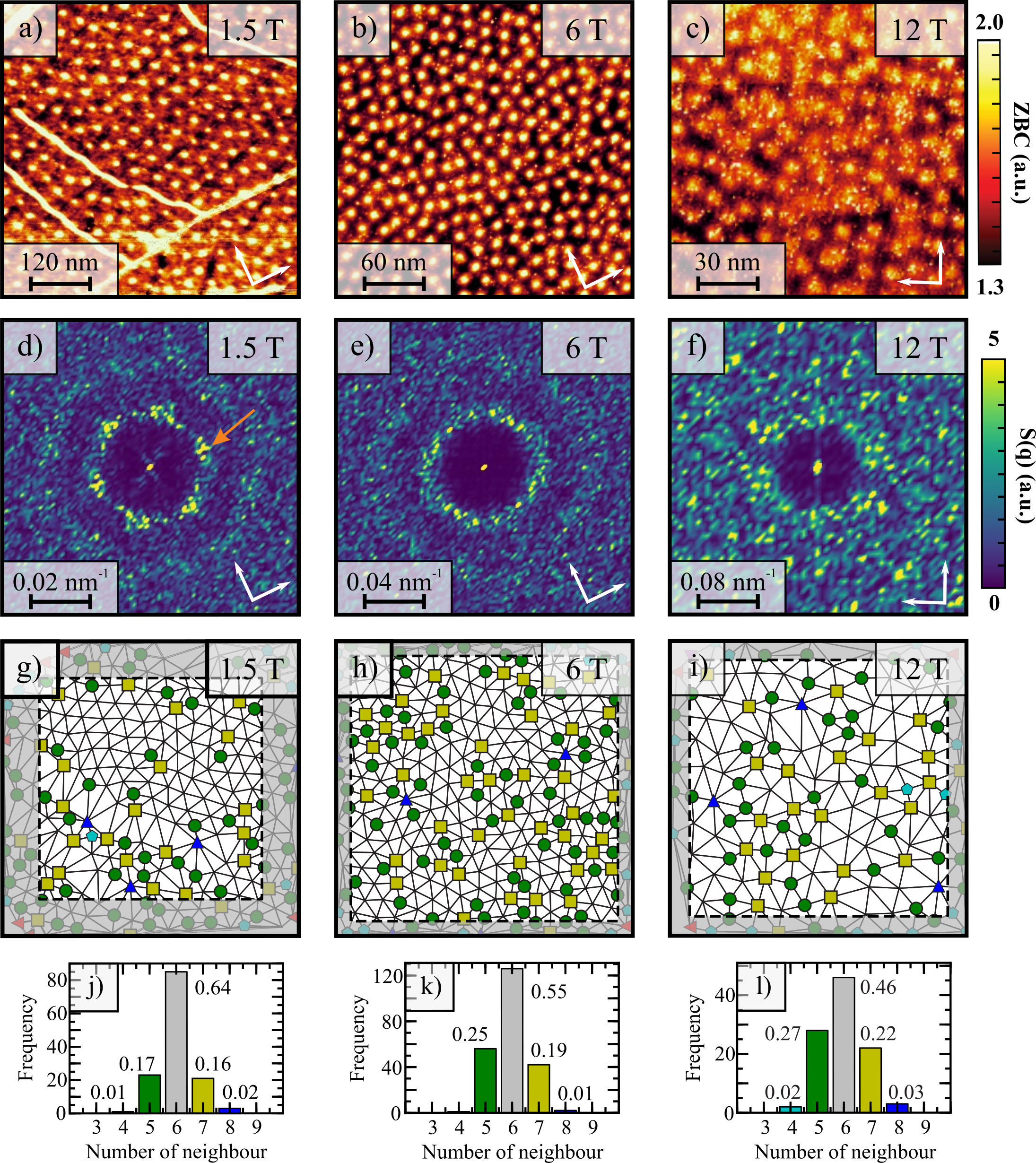}
\caption{Magnetic field dependence of the structural properties of vortex matter nucleated in LiFeAs following a field-cooling process. (a-c) ZBC map revealing vortices at 1.5 T, 6 T and 12 T respectively ($T=5$~K). (d-f) Structure factor $S(q)$ of the vortex positions in (g-i). White arrows mark the orientation of the Fe-Fe-nearest neighbor direction. (g-i) Delaunay-Analysis of the vortex lattice in (a-c). The symbols highlight vortices with less or more than six neighbors: 4: turquoise pentagon, 5: green circle, 7: yellow square, 8: blue triangle. In order to avoid the influence of the lattice defects on the edge of the measuring area, only lattice defects in the non-gray marked area are included in the statistics. (j-l) Statistical analysis of vortex neighbors within the dashed frame in (g-i).}
\label{fig:2_vortices-mag-field1}
\end{figure*}

In order to study the influence of magnetic fields on the vortex lattice, FC lattices of sample 1 in different magnetic fields (1.5 T, 6 T and 12 T) were mapped and analyzed in FIG.~\ref{fig:2_vortices-mag-field1}. For each magnetic field we present the ZBC, revealing the vortex lattice, the corresponding  structure factor $S(q)$ calculated from the vortex positions in (g-i), and a vortex lattice defect characterization which was carried out using the method of Delauney \cite{Delaunay}.  FIG.~\ref{fig:2_vortices-mag-field1}(a) depicts a vortex lattice at 1.5 T, where additional line-defects are visible. $S(q)$ of FIG.~\ref{fig:2_vortices-mag-field1}(a) is shown in (d) and presents a non-closed ring with recognizable diffraction peaks, indicated by the orange arrow.  The formation of the  diffraction peaks confirms a vortex lattice with a certain degree of order. The vortex lattice constant in this case has a value of $a=39.6$~nm.

 An accurate analysis of the vortex lattice defects through Delaunay triangulation is presented in (g). Here the intersection of the connection lines of the vortex position is shown. Usually, in an undisturbed lattice a single vortex has six neighbors. However, lattice perturbations might change the number of neighbors.  In panel (j) we present the statistical distribution of the number of vortex neighbors for each vortex represented by a node in the Delaunay triangulation. The line defects present in FIG.~\ref{fig:2_vortices-mag-field1}(a) do not allow for a clear identification of the vortex core positions in their vicinity due to their high contrast. This leads to a disruption of the Delaunay analysis in those areas. Despite this, we determined the overall defect rate to be 33\%, in which the lattice defects with five and seven neighbors are contributing with 17\% and 16\%, respectively. Defects with higher or lower coordination are negligible. Note that FIG.~\ref{fig:2_vortices-mag-field1}(a) was recorded following a re-cleaving of sample 1, after which the surface was dominated by the observed line defects, previously identified as wrinkles.

FIG.~\ref{fig:2_vortices-mag-field1}(b) shows a vortex lattice without line defects taken at 6 T. Apparently, the lattice has no recognizable order at this field value. This is confirmed by the absence of clear Bragg-peaks in $S(q)$ (FIG.~\ref{fig:2_vortices-mag-field1}(e)). Indeed, the ring in $S(q)$ indicates a vortex glass configuration without long range order and a vortex separation of $a = 19.9$~nm \cite{Blatter1997}. The lattice defect characterization is shown in FIG.~\ref{fig:2_vortices-mag-field1}(h) and (k). It is seen that the defect rate of 44 \% is now higher. Defects with 5 and 7 neighbors are dominant, however, with an increment of  5 neighbor defects.

Finally, for the highest magnetic field (12~T), FIG.~\ref{fig:2_vortices-mag-field1}(c) shows the appearance of a highly distorted vortex lattice, where the contrast between the vortices and the superconducting area is not completely clear. The corresponding $S(q)$, which is shown in FIG.~\ref{fig:2_vortices-mag-field1}(f), reflects a diffuse circular shape with a corresponding lattice constant of roughly $a \approx 12.2 $~nm. The defect analysis in FIG.~\ref{fig:2_vortices-mag-field1}(i,l) shows an expected increase of the defects abundance to a value of 49\% where now 5 and 7 neighbor defects dominate with an increment of the 4 and 8 neighbor defects.   

\begin{figure*}
\includegraphics[scale=0.6]{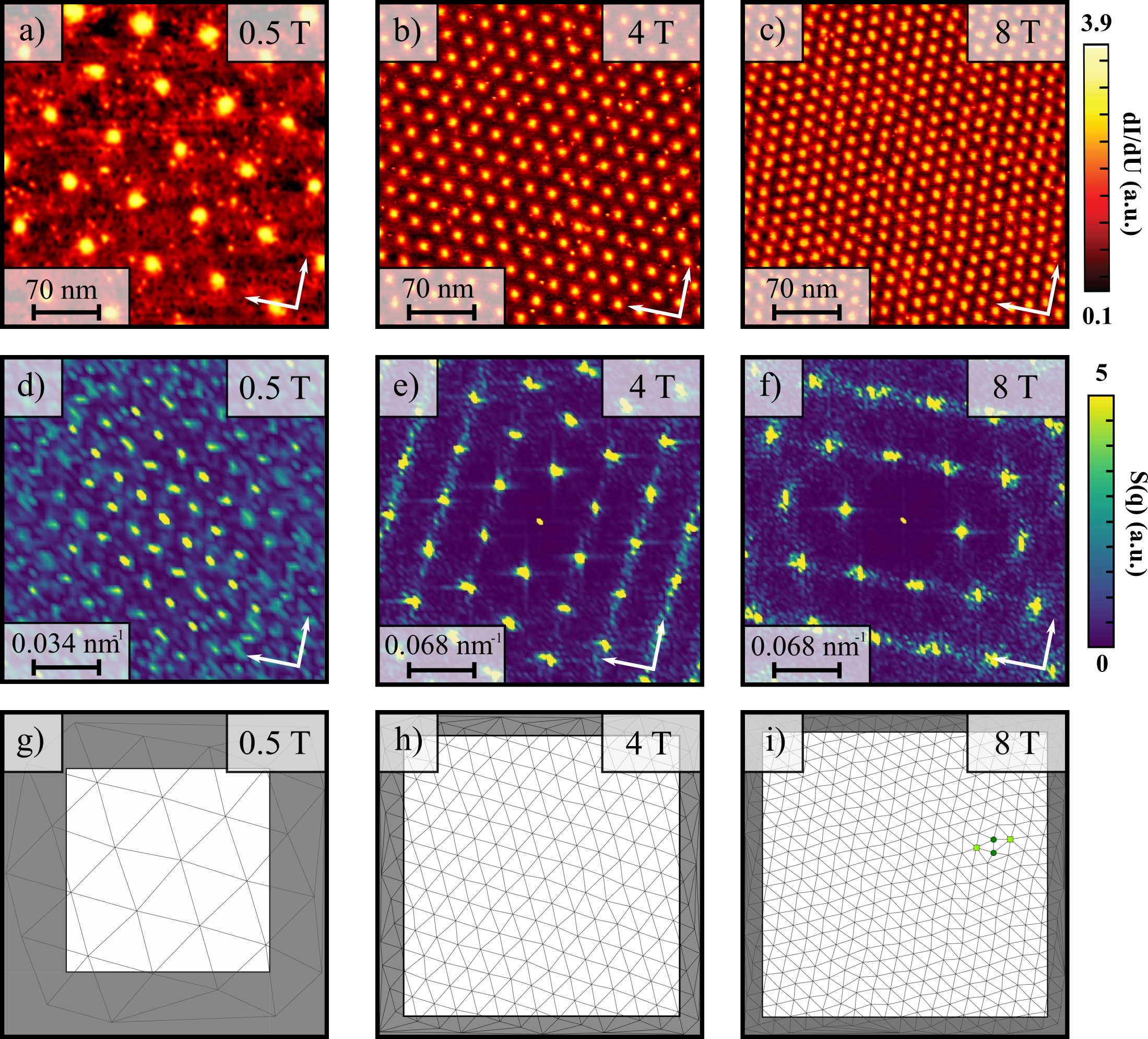}
\caption{Magnetic field dependence of the structural properties of vortex matter nucleated in LiFeAs following a zero-field-cooling process. (a-c) dI/dU maps revealing vortices at 0.5 T, 4 T and 8 T respectively ($U=-3$~mV , $I=100$~pA, $T=6$~K). (d-f) Structure factor $S(q)$ of the vortex positions in (g-i). White arrows mark the orientation of the Fe-Fe-nearest neighbor direction. (g-i) Delaunay-Analysis of the vortex lattice in (a-c). The symbols show vortices with less or more than six neighbors: 5: green circle, 7: light green square.}
\label{fig:3_vortices-mag-field2}
\end{figure*}

FIG.~\ref{fig:3_vortices-mag-field2}(a-c)  shows dI/dU maps of sample 2 recorded under ZFC conditions at 6~K and at magnetic fields of 0.5~T, 4~T and 8~T, respectively. The vortex lattice in sample 2 shows a much higher degree of order than the one of sample 1, in accordance with previous reports for samples prepared under ZFC conditions \cite{Bhattacharya1993,Banerjee1999,Zhang2019a}. The triangular lattice geometry is clearly identifiable even at 8 T. This is even better seen in the $S(q)$ images in FIG.~\ref{fig:3_vortices-mag-field2}(d-f). Here clear diffraction peaks of a triangular lattice can be seen at all fields.  The Delaunay analysis, depicted in FIG.~\ref{fig:3_vortices-mag-field2}(g-i), makes evident that there are differences between field-cooled and zero-field-cooled vortex structures, with the ZFC structure being more ordered. For 0.5 T and 4 T no lattice defects could be found and for 8 T the total number of lattice defects amounts to 4, resulting in a defect abundance of only 1 \%. The lattice constant of the vortex lattices extracted from the structure factor for each field are   $a_{0.5T}$ = 72.2 nm, $a_{4T}$ = 24.9 nm and $a_{8T}$ = 18.2 nm, respectively. Furthermore, no square lattice transition can be observed at 8 T for vortex lattices nucleated in ZFC conditions.

\section{Discussion}

The Ginzburg-Landau coherence length values $\xi_{GL}^{(1)}=(4.4~\pm0.5)$~nm and $\xi_{GL}^{(2)}=(4.1~\pm0.5)$~nm of sample 1 and sample 2 agree within error bars. We estimate their respective upper critical fields via $H_{c2} = \frac{\phi_0}{2\pi\xi_{GL}^2}$ and obtain $H_{c2} = (17~\pm~3)$~T and $H_{c2}= (19.6~\pm~4)$~T, respectively. These values are in good agreement with experimental findings of other groups \cite{Song2010,Zhang2011,Cho2011,Khim2011,Inosov2010,Kurita2011,Lee2009,Li2013,Heyer2011}.

Despite this consistency of the measured $\xi_{GL}$ and $H_{c2}$ between the samples, there is an obvious strong difference in the degree of vortex lattice order for both cooling processes. This behavior can naturally be explained by the fast flow of flux-lines into the superconductor from its edges upon ramping the field from zero to a finite value in ZFC conditions. This drastically enhances vortex-vortex interactions in respect to pinning effects, allowing the vortex matter to settle in configurations with higher degrees of order \cite{Bhattacharya1993,Banerjee1999}. 

Our findings of a highly ordered vortex lattice following a ZFC process at first glance is in good agreement with findings by \textit{Hanaguri et al.} \cite{Hanaguri2012} and \textit{Zhang et al.} \cite{Zhang2019a}. However, there is a surprising difference with respect to the $C_6\rightarrow C_4$ symmetry transition: While \textit{Hanaguri et al.} observes a transition at around 8 T in FC conditions, \textit{Zhang et al.} reports this transition to occur at 3-4~T \cite{Zhang2019a} for the ZFC case. The vortex lattice of both of our samples, irrespective of the cooling process, appears to remain $C_6$ symmetric even at the maximum field measured of 12~T (see FIG.~\ref{fig:2_vortices-mag-field1}(c,f,i,l)). Furthermore, also different from the findings of $Zhang~et~al$., the $C_6$ vortex lattice of sample 2 is locked to the crystal lattice at all fields (see FIG.~\ref{fig:3_vortices-mag-field2}(d,e,f)), while $Zhang~et~al.$ report such a locking only in the $C_4$ high field phase. It was argued \cite{Zhang2019a}, that the $C_6\rightarrow C_4$ lattice transition occurs once a sufficient overlapping of the vortex bound states is realized which $Zhang~et~al.$ estimate at the intervortex distance $ a\approx 5-6~\xi_{GL}$. The absence of this transition in our samples, which have the same value of $\xi_{GL}$ within the error range as the one investigated by $Zhang~et~al.$, suggests that it is not primarily the overall sample independent Ginzburg-Landau coherence length or the nature of the cooling process which determines this transition to occur. Therefore, other sample-dependent properties rather seem to play a role here.

\textit{Hanaguri et al.} conjectured that, in analogy to previous observation on other high-T$_c$-superconductors \cite{Gilardi2002,Curran2011,Ray2014,Kogan1997,Sakata2000,Nishimori2004}, vortex lattice symmetry in LiFeAs is affected by the anisotropy of the underlying superconducting order parameter. In this context it is interesting to note that a number of recent experimental and theoretical works have suggested the possibility of multiple superconducting order parameters existing in LiFeAs \cite{Baek2012,Baek2013,Cao2021,Nag2016,Ahn2014}. A change to the order parameter, possibly induced by slight differences in sample stoichiometry, could explain the observed contrasting behavior of vortex matter in LiFeAs. Furthermore, recent observations of nematic ordering in LiFeAs \cite{Kushnirenko2020} provide another natural explanation for a symmetry reduction in the superconducting state. Another possible explanation is that in the samples studied in Refs.~\onlinecite{Hanaguri2012,Zhang2019a} the magnitude of the magneto-elastic effect might be larger than in our samples. The magneto-elastic effect is expected to induce the symmetry transition in tetragonal superconductors \cite{Lin2017} and in the case of LiFeAs it has a moderate magnitude, proportional to $(dT_{c}/dP)^2$, since in LiFeAs the derivative is of the order of 1\,K/GPa \cite{Gooch2009}.


Given the great attention paid on the possibility of topological superconductivity in LiFeAs \cite{Zhang2019,Kong2021,Liu2021}, it thus seems worthwhile to systematically study the the influence of the sample to the superconducting properties in future work.


\begin{acknowledgments}

This work received support from the Deutsche Forschungsgemeinschaft through the Priority Programme SPP1458 (Grant HE3439/11 and BU887/15-1), and the Graduate School GRK1621. Furthermore, this project received funding from the European Research Council (ERC) under the European Unions Horizon 2020 research and innovation program (grant agreement No 647276 – MARS – ERC-2014-CoG). Y.F. acknowledges funding from the Georg Forster research prize from the Alexander von Humboldt Stiftung. S.S. acknowledges funding by the Deutsche Forschungsgemeinschaft via the Emmy Noether Programme ME4844/1-1 (project id 327807255)

\end{acknowledgments}

\appendix
\section{}

\begin{figure*}
\includegraphics[scale=0.6]{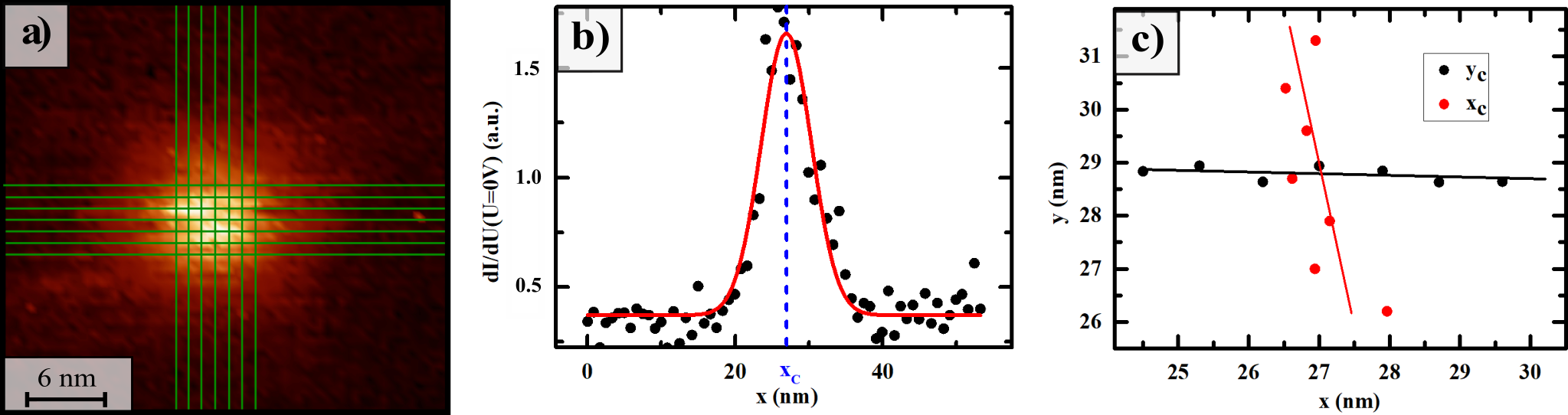}
\caption{a) ZBC map of a vortex core of sample 2. Profiles were extracted along the green lines. Panel b) shows an example profile of one line in a) as well as the corresponding Gaussian fit. c) $x_c(y)$ and $y_c(x)$ as a result of the fits. The crossing of the lines marks the center of the vortex.  }
\label{fig:SingleVortexsample2Center}
\end{figure*}

\begin{figure}
\includegraphics[scale=0.35]{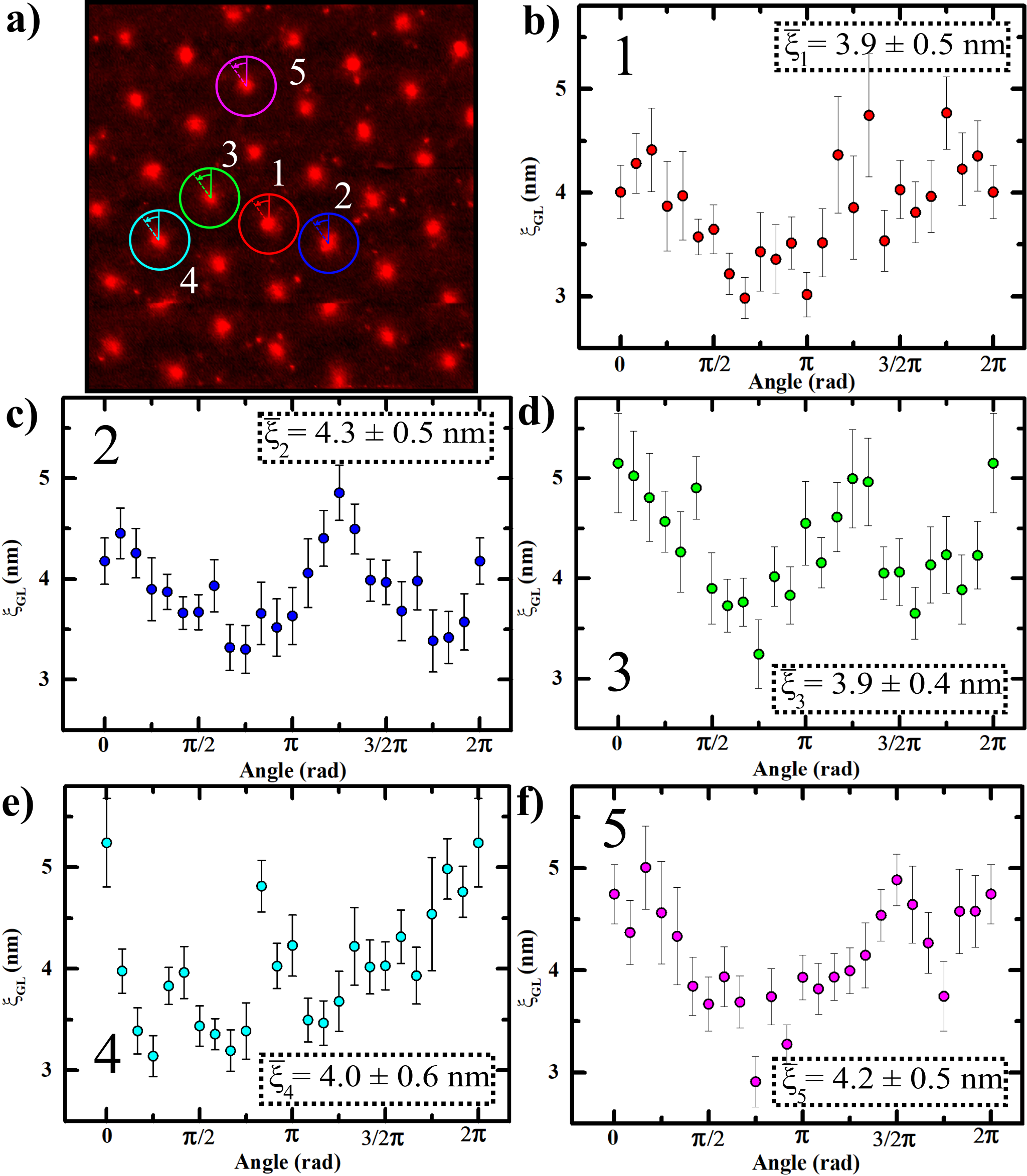}
\caption{a) 210x210 nm ZBC map of vortex lattice for sample 2 under ZFC conditions ($B=2$~T, $T=$~6K).  $\xi_{GL}$ was calculated for 5 vortices as marked in the image. Panels  b)-f) show the individual results of $\xi_{GL}(\alpha)$ as well as the mean value  $\bar{\xi_{GL}}$. }
\label{fig:SingleVortexsample2Map}
\end{figure}

It is important for an accurate determination of $\xi$, using the method described in this publication, to find the exact center of the analyzed vortex cores. To achieve this aim, we used Gaussian fits of profiles from the ZBC maps of the individual vortices. Multiple profiles, parallel to the x and y axis were extracted and from this information the center point $x_c(y)$ and $y_c(x)$ could be determined. By plotting the resulting values together and performing linear interpolation we obtain two lines which intersect at the vortex center. This method is shown in FIG.~\ref{fig:SingleVortexsample2Center} for a vortex core of sample 2. In order to get a more robust result for the GL-coherence length of sample 2, we analyzed multiple vortex cores from the ZBC map shown in FIG.~\ref{fig:SingleVortexsample2Map}. The 5 vortices marked were chosen because their surroundings appear free of obvious defects (visible as bright spots in the image). The results for the individual vorticies are shown in Panels b)-f), where a mean value of $\xi_{GL}^{2} = 4.1 \pm 0.5$ nm was calculated.

\begin{figure}
\includegraphics[scale=0.3]{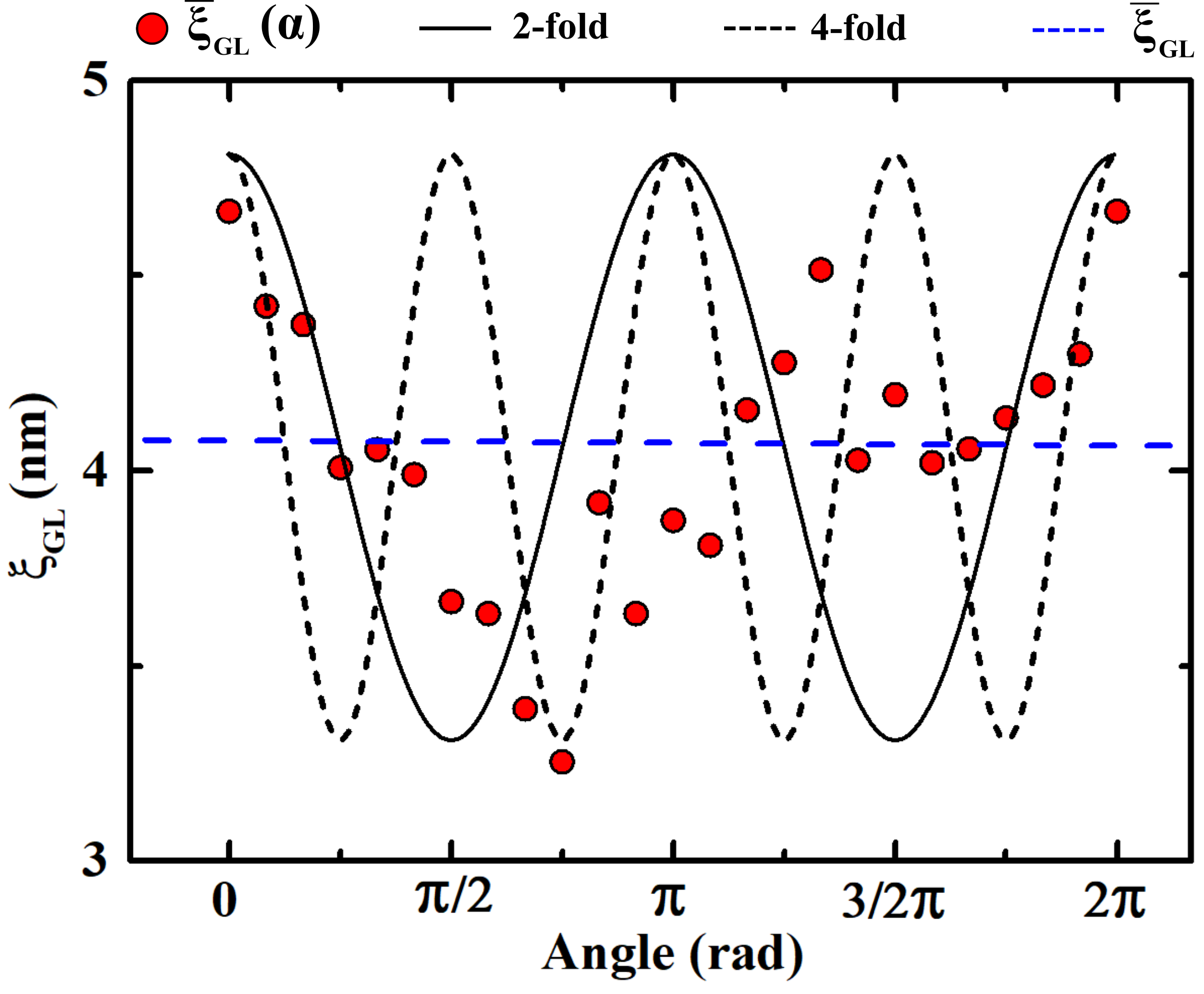}
\caption{Angle dependent average $\bar{\xi}_{GL}(\alpha)$ for all 5 vortex cores from FIG.~\ref{fig:SingleVortexsample2Map}. The solid and dashed black curves are of the form $\propto cos(2\alpha)$ and $\propto cos(4\alpha)$, representing a 2-fold and 4-fold symmetry respectively. The dashed blue line marks the mean value of $\xi_{GL}$.}
\label{fig:SingleVortexsample2Map_Mean}
\end{figure}

FIG.\ref{fig:SingleVortexsample2Map_Mean} shows the average value $\bar{\xi}_{GL}(\alpha)$ from all five vortex cores analyzed in FIG.~\ref{fig:SingleVortexsample2Map}, revealing some anisotropy. This makes it unlikely that defects are the sole reason for the observed deviations of $\xi_{GL}(\alpha)$. Nevertheless, FIG.~\ref{fig:SingleVortexsample2Map_Mean}  further demonstrates that the observed anisotropy is of low symmetry. Very clearly, this low-symmetric angle dependence is neither compatible with the  4-fold symmetry which is reported for LiFeAs \cite{Hanaguri2012,Wang333} nor with a thinkable 2-fold symmetry which would be expected for coupling to a nematic order parameter \cite{Lu2018,Putilov2019}. This is visualized by the dashed solid black curves representing the expected distribution of $\xi_{GL}(\alpha)$ for generic 4-fold and 2-fold symmetries, respectively.  Note that a possible explanation for this observation might be the influence of drift on the measurements. Here, drift could deform an otherwise isotropic vortex, creating the observed anisotropy, or distort a possible intrinsic anisotropy of the vortex cores to reduce its symmetry. We therefore refrain from drawing any further conclusions from this observation.

\begin{figure}
\includegraphics[scale=0.6]{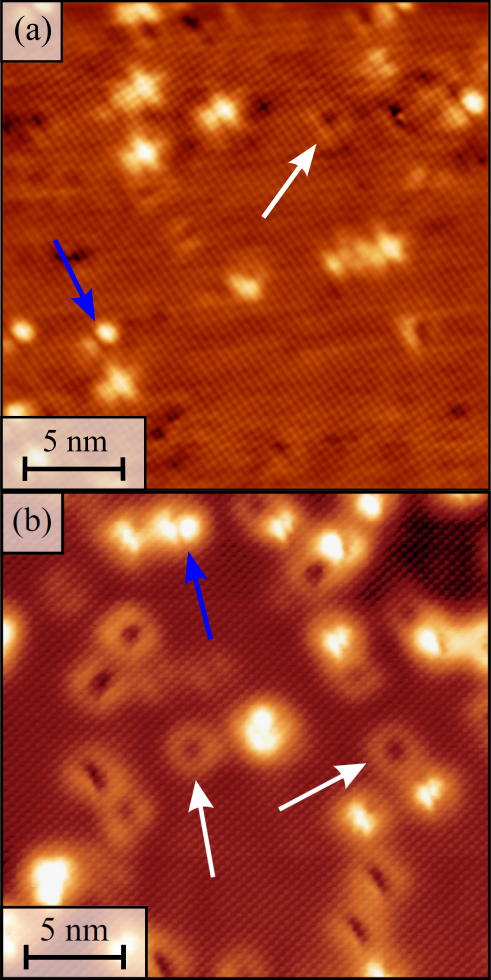}
\caption{(a)Topography image (25x25 nm, $U_{bias} = 30$ mV, $I = 300$ pA) of the surface of sample 2. (b) Topography image (25x25 nm, $U_{bias} = 35$ mV, $I =$ 300 pA) of the surface of sample 1 from \cite{Schlegel2014}. Both topographies show the atomic corrugation of LiFeAs as well as some of the commonly found intrinsic defects. White and blue arrows mark the location of As-D$_4$ and Fe-D$_4$ defects, respectively \cite{Grothe2012}. }
\label{fig:AR}
\end{figure}

FIG.~\ref{fig:AR} shows atomically resolved images of samples 1 and 2. The intrinsic atomic defects \cite{Grothe2012,Schlegel2017} commonly observed for LiFeAs are visible. From surface topography of this kind we can roughly estimate that the surface defect concentration in both samples is below 0.5 \% per unit cell, speaking for the high quality of our samples. However, determining the precise bulk stoichiometry of our samples based on a limited number of surface topographies with a generally small field of view (FOV) turns out to be difficult and unreliable.

\newpage

\thispagestyle{empty}

\newpage
\bibliography{Absence_of_hexagonal_to_square_structural_transition_in_LiFeAs_vortex_matter}

\end{document}